\documentclass[twocolumn,showpacs]{revtex4}

\usepackage{graphicx}%
\usepackage{dcolumn}
\usepackage{amsmath}

\makeatletter
\def\btt#1{\texttt{\@backslashchar#1}}%
\DeclareRobustCommand\bblash{\btt{\@backslashchar}}%
\makeatother

\begin{document}

\title[Short Title]{Higher Dimensional Radiation
 Collapse and Cosmic Censorship}

\author{S. G. Ghosh}
\email{sgghosh@yahoo.com}
\affiliation{%
Department of Mathematics, Science College, Congress Nagar,
Nagpur-440 012, INDIA}%

\author{R. V.~Sararykar}
\email{sarayaka@nagpur.dot.net.in} \affiliation{Department of
Mathematics,
Nagpur University,  Nagpur-440 010, INDIA}%

\date{\today}

\begin{abstract}
We study the occurrence of naked singularities in the spherically
symmetric collapse of radiation shells  in a higher dimensional
spacetime. The necessary conditions for the formation of a naked
singularity or a black hole are obtained. The naked singularities
are found to be strong in the Tipler's sense and thus  violating
cosmic censorship conjecture.
\end{abstract}


\maketitle

The cosmic censorship conjecture (CCC) put forward some three decades ago
by Penrose \cite{rp} says that in generic situation all singularities
arising from regular initial data are clothed by event horizon.
Since CCC remains as an unresolved issue in classical general relativity,
examples which appear to violate conjecture remain important and may be
valuable when one attempts to formulate notion of CCC in concrete mathematical
form.  The Vaidya
solution  \cite{pc} is most commonly used as a testing ground for various
forms of the CCC.  In particular, Papapetrou \cite{pp} first showed that this
solution can give rise to the formation of naked singularities and thus
provided one of the earlier counter examples to CCC.  Since then,
the solution is extensively studied in gravitational
collapse with reference to CCC (see e.g. \cite{pj} and references therein).
Lately, there has been interest in studying gravitational collapse in higher
dimensions \cite{hd}.   This brief report searches for the occurrence of naked
singularities in higher dimensional Vaidya spacetime and if they do,
to investigate
whether the dimensionality of spacetime has any role in the nature of
singularities.
We find that higher dimensional Vaidya spacetime admit
strong-curvature naked
singularities in the Tipler's \cite{ft} sense.

The idea that spacetime should be extended from four to higher dimensions
was introduced
by Kaluza and Klein \cite {kk} to unify gravity and electromagnetism.  Five
dimensional $(5D)$ spacetime is particularly more relevant because both
$10D$ and $11D$ super-gravity theories yield solutions where a $5D$ spacetime
results after dimensional reduction \cite{js}.  Hence,  we shall confine
 ourselves to $5D$
case.  It should be noted that higher dimensional spacetimes
where all dimensions are in the equal foot, like ones to be
considered here, are not so realistic, as we are living in
effectively 4-dimensional spacetime.  So in principle one might
expect that by dimensional reduction the higher dimensional
spacetime should reduce to our 4-dimensional world.  In this
sense, the models considered in this paper are ideal.

The metric of collapsing Vaidya models in $5D$ case is \cite{iv}
\begin{equation}
ds^2 = - (1 -  \frac{m(v)}{r^2}) dv^2 + 2 dv dr + r^2 d \Omega^2
\label{eq:me}
\end{equation}
where $v$ is null coordinate which represents advanced Eddington time with
  $-\infty < v < \infty$, $r$ is radial coordinate with
$0\leq r < \infty$, $ d\Omega^2 = d \theta_1^2+ sin^2 \theta_1 (d
\theta_2^2+sin^2 \theta_2 d\theta_3^2)$ is a metric of 3 sphere
and the arbitrary function $m(v)$ (which is restricted only by
the energy conditions), represents the mass at advanced time $v$.
The energy momentum tensor can be written in the form
\begin{equation}
 T_{ab} = \frac{3}{2 r^3} \dot{m}(v) k_{a}k_{b} \label{eq:tn}
\end{equation}
with the null vector $k_{a}$ satisfying $k_{a} = -
\delta_{a}^{v}$  and $ k_{a}k^{a} = 0$. We have used the units
which fix the speed of light and gravitational constant via $8
\pi G\;=\;c=\;1.$ Clearly, for the weak energy condition to be
satisfied we require that $\dot{m}(v)$ to be non negative, where
an over-dot represents a derivative with respect to $v$. Thus
mass function is a non-negative increasing function of $v$ and
that radiation is imploding.

The physical situation here is that of a radial influx of null fluid
in an initially empty region of the higher dimensional
Minkowskian spacetime.
The first shell
arrives at $r=0$ at time $v=0$ and the final at $v=T$.
A central singularity of growing mass is
developed at $r=0$.
  For $ v < 0$ we have $m(v)\;=\;0$, i.e., higher dimensional
Minkowskian spacetime,
 and for $ v > T$,
$\dot{m}(v)\;=\;0$, $m(v)\;$ is positive
definite.  The metric for $v=0$ to $v=T$ is higher dimensional
Vaidya, and for $v>T$ we have the higher dimensional Schwarzschild
solution.
In order to get an analytical solution for our higher dimensional case,
we choose $m(v) \propto v^2$. To be specific we take
\begin{equation}
m(v) = \left\{ \begin{array}{ll}
    0,          &   \mbox{$ v < 0$}, \\
    \lambda v^2 (\lambda>0) &   \mbox{$0 \leq v \leq T$}, \\
    m_{0}(>0)       &   \mbox{$v >  T$}.
        \end{array}
    \right.             \label{eq:mv}
\end{equation}
and
with this choice of $m(v)$ the spacetime is self similar \cite{ss},
admitting a homothetic Killing vector $\xi^a$ given by
\begin{equation}
\xi^a = r \frac{\partial}{\partial r}+v \frac{\partial}{\partial v} \label
{eq:kl1}
\end{equation}
which satisfies the condition
\begin{equation}
\L_{\xi}g_{ab} = \xi_{a;b}+\xi_{b;a} = 2g_{ab}
\end{equation}
, $\L$ denotes the Lie
derivative.  It follows that
$\xi^a k_{a}$
is constant along radial null geodesics and hence a constant of motion,
\begin{equation}
\xi^a K_{a} = r K_r + v K_v = C \label{eq:kl2}
\end{equation}

Let $K^{a} = \frac{dx^a}{dk}$ be the tangent vector to the null geodesics,
where $k$ is an affine
parameter.   Geodesic equations, on
using the null condition $K^{a}K_{a} = 0$, take the simple form
\begin{equation}
\frac{dK^v}{dk} + \frac{m(v)}{r^3} (K^v)^2 = 0  \label{eq:kv1}
\end{equation}
\begin{equation}
\frac{dK^r}{dk} + \frac{\dot{m}(v)}{2 r^2} (K^v)^2 = 0  \label{eq:kr1}
\end{equation}
Following \cite{dj,gb}, we introduce
\begin{equation}
K^v = \frac{P}{r}   \label{eq:kv}
\end{equation}
and, from the null condition, we obtain
\begin{equation}
K^r = \left(1 - \frac{m(v)}{r^2} \right) \frac{P}{2r}   \label{eq:kr}
\end{equation}
The function $P(v,r)$ obeys the differential equation
\begin{equation}
\frac{dP}{dk} - \left(1 - \frac{3 m(v)}{r^2} \right) \frac{P^2}{2 r^2} = 0
\label{eq:pde}
\end{equation}
Radial  null
geodesics of the metric (\ref{eq:me}), by virtue of Eqs. (\ref{eq:kv})
and  (\ref{eq:kr}), satisfy
\begin{equation}
\frac{dr}{dv} = \frac{1}{2} \left[1 -  \frac{m(v)}{r^2}  \right]
\label{eq:de1}
\end{equation}
Clearly, the above differential equation has a singularity at $r=0$, $v=0$.
The nature (a naked singularity or a black hole) of the collapsing solutions
can be characterized by the existence of radial null geodesics coming out from
 the singularity.
Eq. (\ref{eq:de1}), upon using  eq. (\ref{eq:mv}), turns out to be
\begin{equation}
\frac{dr}{dv} = \frac{1}{2} \left[1 - \lambda X^2  \right]
\label{eq:de2}
\end{equation}
where $X \equiv v/r$ is the tangent to a possible outgoing geodesic.
In order to determine the nature of the limiting value of $X$ at $r=0$, $v=0$
on a singular geodesic, we let
\begin{equation}
X_{0} = \lim_{r \rightarrow 0 \; v\rightarrow 0} X =
\lim_{r\rightarrow 0 \; v\rightarrow 0} \frac{v}{r}      \label{eq:lm1}
\end{equation}
Using (\ref{eq:de2}) and L'H\^{o}pital's rule we get
\begin{equation}
X_{0} = \lim_{r\rightarrow 0 \; v\rightarrow 0} X =
\lim_{r\rightarrow 0 \; v\rightarrow 0} \frac{v}{r}=
\lim_{r\rightarrow 0 \; v\rightarrow 0} \frac{dv}{dr} =
\frac{2}{1 -  \lambda X_{0}^2}  \label{eq:lm2}
\end{equation}
which implies,
\begin{equation}
\lambda X_{0}^3 - X_{0} + 2 = 0    \label{eq:ae}
\end{equation}

This algebraic equation governs the behavior of the tangent vector near the
 singular point.  Thus by studying the solution of this algebraic equation,
the nature of the singularity can be determined.  The central shell focusing
    singularity is at least locally naked (for brevity we have addressed it as
naked throughout this paper), if eq. (\ref{eq:ae})
 admits one or more positive
real roots.
When there are no positive real roots to eq. (\ref{eq:ae}),
the central
singularity is not naked because in that case there are no outgoing future
directed null geodesics from the singularity.  Hence in the absence of
positive real roots, the collapse will always lead to a black hole.
The condition under which this locally naked singularity could be globally
naked is well discussed \cite{pj} and we shall not discuss it here.
Thus, the
occurrence of positive real roots implies that the strong CCC is violated,
though
 not necessarily the weak CCC.  We now examine the condition for the
occurrence of a naked singularity.  Eq. (\ref{eq:ae}) has two
positive roots if $\lambda \leq 1/27$, e.g., the two positive
roots of eq. (\ref{eq:ae})
  $X_{0}= 2.21833$ and $5.69593$,
correspond to $\lambda = 1/50$.  Thus referring to above discussion,
gravitational collapse of null fluid in higher dimensions leads to a naked
singularity if $\lambda \leq 1/27$ and to formation of a black hole
otherwise.  The degree of inhomogeneity of collapse is defined as
$\mu \equiv 1/ \lambda$ (see \cite{jl}), we see that for a collapse
 sufficiently inhomogeneous, naked singularities develop.  Comparison with
analogous $4D$ case shows that naked singularity occurs for a
slightly larger value of inhomogeneity factor in $5D$.

The analysis of geodesics escaping to far away observers emanating
from singularity is similar to that discussed in \cite{pp,dj}. To
see this, we first note that the apparent horizon in our case is
given by $X=1/ \sqrt{\lambda}$.  We write the equation of
geodesics in the form $r = r(X)$.  Using eqs. (\ref{eq:mv}) and
(\ref{eq:de2}) we obtain,
\begin{equation}
r \frac{dX}{dr} = \frac{\lambda X^3 - X + 2}{1 - \lambda X^2}
\label{eq:rx}
\end{equation}
Integration of the above equation yields the equation of geodesics
(integral curves). Here, we briefly describe constraints when
geodesics can escape to a far away observer. Basically this
follows from  analysis of eq. (\ref{eq:rx}) and its solution. We
consider separately the following two cases.

\noindent {Case (a)} $\lambda< 1/27$

The solution of eq. (\ref{eq:rx}) is
\begin{equation}
r = \frac{(x-a_1)^{C_1}}{(x-a_2)^{C_2}(x+a_3)^{C_3}}
\label{eq:rxs}
\end{equation}
where $a_1$ and $a_2$ are two positive roots with $a_1 > a_2$ and
$-a_3$ is the negative root of equation eq. (\ref{eq:ae}). $C_1$,
$C_2$ and $C_3$ are given by
\begin{eqnarray*}
C_1 = && \frac{a_2 a_3}{(a_1-a_2)(a_1+a_3)}, \hspace{.1in} C_2 =
\frac{a_1 a_3}{(a_1-a_2)(a_2+a_3)}, \\ \nonumber
&& C_3 =
\frac{a_1 a_2}{(a_1+a_3)(a_2+a_3)}
\end{eqnarray*}
For $X \geq 1/ \sqrt{\lambda}$, all geodesics are ingoing and no
geodesics escapes.  For $X < 1/ \sqrt{\lambda}$,  geodesic
families can meet the singularity with tangent $X=a_1$ and
$r=\infty$ can be realized in future along the same geodesic at
$X=a_2$. Thus the singularity will be globally naked and an
infinity of geodesics would escape from singularity to reach far
away observer.

It is also seen that for very small value of $\lambda$, say
$O(10^{-17})$, $X \geq 1/ \sqrt{\lambda}$ and hence singularity is
only locally naked.

\noindent Case (b) $\lambda=1/27$

In this case, the eq. (\ref{eq:rx}) takes the form
\begin{equation}
r \frac{dX}{dr} = \frac{(X-3)^2 (X+6)}{27-X^2}
\end{equation}
which has a trivial solution $X=3$. In the variables (t, r), it
takes the form $t=2 r$ which is a special null geodesic.  In
analogous $4D$ case one gets $t=3r$ \cite{pp}. For $X \neq 3$, eq.
(\ref{eq:rx}) can be integrated to
\begin{equation}
  r = \frac{Exp \left( {\frac{2}{3-X}} \right) }{(3-X)^{8/9} (X+6)^{1/9}}
  \label{eq:rxs1}
\end{equation}
As in $4D$ case \cite{pp}, we can conclude that there is an
infinite number of geodesics escaping to a far away observer.

 The strength of singularity is an important issue
because there have been attempts to relate it to stability
\cite{djd}. A singularity is termed gravitationally strong or
simply strong, if it destroys by crushing or stretching any
object which falls in to it.  A sufficient condition \cite{ck}
for a strong singularity as defined by Tipler \cite{ft} is that
for at least one non-space like geodesic with affine parameter
$k$, in limiting approach to singularity, we must have
\begin{equation}
\lim_{k\rightarrow 0}k^2 \psi =
\lim_{k\rightarrow 0}k^2 R_{ab} K^{a}K^{b} > 0 \label{eq:sc}
\end{equation}
where $R_{ab}$ is the Ricci tensor. Eq. (\ref{eq:sc}), with the help of
eqs. (\ref{eq:tn}), (\ref{eq:mv}) and (\ref{eq:kv}), can be expressed as
\begin{equation}
\lim_{k\rightarrow 0}k^2 \psi =
\lim_{k\rightarrow 0} 3 \lambda X (\frac{kP}{r^2})^2 \label{eq:sc1}
\end{equation}
Our purpose here is to investigate the above condition along
future directed null geodesics coming out from the singularity.
For this, solution of eq. (\ref{eq:pde}) is required. Eq.
(\ref{eq:kl2}), because of eqs. (\ref{eq:mv}), (\ref{eq:kv}) and
(\ref{eq:kr}), yields
\begin{equation}
P = \frac{2 C}{2- X +\lambda X^3} \label{eq:ps}
\end{equation}
which is a solution of eq. (\ref{eq:pde}) and geodesics are completely
determined. Further, we note that
\begin{equation}
\frac{dX}{dk} = \frac{1}{r} K^v - \frac{X}{r} K^r \label{eq:xk}
\end{equation}
which, on inserting the expressions for $K^v$ and $K^r$, become
\begin{equation}
\frac{dX}{dk} = (2 - X - \lambda X^3) \frac{P}{2r^2} =
\frac{C}{r^2}
 \label{eq:xk1}
\end{equation}
Using the fact that as singularity is approached, $k \rightarrow 0$,
$r \rightarrow 0$ and  $X \rightarrow a_{+}$ (a root of  (\ref{eq:ae})) and
using  L'H\^{o}pital's rule, we observe
\begin{equation}
\lim_{k\rightarrow 0} \frac{kP}{r^2} = \frac{2}{1+\lambda X_{0}^2}
\end{equation}
and hence eq. (\ref{eq:sc1}) gives
\begin{equation}
\lim_{k\rightarrow 0}k^2 \psi =
\frac{12 \lambda X_{0}}{(1+\lambda X_{0}^2)^2} >0
\end{equation}
Thus along radial null geodesics strong curvature condition is satisfied.

Recently, Nolan \cite{bc} gave an alternative approach to check
the nature of singularities without having to integrate the
geodesic equations.  It was shown in \cite{bc} that
 a radial null geodesic which runs into $r=0$ terminates in a  gravitationally
weak singularity if and only if $\dot{r}$ is finite in the limit as the
singularity is approached (this occurs at $k=0$), the over-dot here indicates
differentiation along the geodesics.  So assuming a weak singularity, we have
\begin{equation}
\dot{r} \sim  d_{0} \hspace{0.2in} r \sim  d_{0} k
\end{equation}
Using the asymptotic relationship above and the form of $m(v)$, the geodesic
equations yield
\begin{equation}
\frac{d^2v}{dk^2} \sim   \alpha  k^{-1}
\end{equation}
where $\alpha = - \lambda X_{0}^4 d_{0} =$ a non zero const.,
which is inconsistent with
 $\dot{v} \sim  d_{0}X_{0}$, which is finite.  Since,
the coefficient $\alpha$ of
$k^{-1}$ is non-zero, the singularity is gravitationally strong.
Having seen that naked
singularity is a strong curvature singularity, we now turn our attention to
check it for scalar polynomial singularity.  We therefore examine the behavior
 of Kretschmann scalar ($K = R_{abcd} R^{abcd}$, $R_{abcd}$
is the Riemann tensor). The Kretschmann scalar with the help of
eq. (\ref{eq:mv}), takes the form
\begin{equation}
K = \frac{72}{r^4} X^4  \label{eq:ks1}
\end{equation}
which diverges at the naked singularity and hence the singularity is a scalar
polynomial singularity.
The Weyl scalar ($C = C_{abcd} C^{abcd}$, $C_{abcd}$ is the Weyl tensor) has
the same expression as Kretschmann scalar and
thus the Weyl scalar also diverges at the naked singularity and so the
singularity is physically significant \cite{bs}.

The Vaidya metric in $4D$ case has been extensively used to study
the formation of naked singularity in spherical gravitational
collapse \cite{pj}. We have extended this study to higher
dimensional Vaidya metric, and found that strong curvature naked
singularities  do arise for slightly higher value of inhomogeneity
parameter. We have checked, for naked singularities to be
gravitationally strong, by method in \cite{ck} and by alternative
approach proposed by Nolan \cite{bc} as well, and both seem to be
in agreement. It is straight forward to extend above analysis for
non radial causal curves. Further, the Kretschmann and Weyl
scalars blows up as singularity is approached. In conclusion, this
offers a counter example to CCC.\\

 Authors
would like to thank IUCAA, Pune (INDIA) for the hospitality.
Thanks are also due to Brien C. Nolan
 for helpful correspondence.

\noindent

\begin{references}
\bibitem{ss} A spherical symmetric space-time is self similar if
$ g_{tt} (ct, cr) = g_{tt}(t,r)$ and  $ g_{rr} (ct, cr) = g_{rr}(t,r)$ for
every $c>0$.
\end{references}
\end{document}